# Vibrational Properties of the Bi$_2$Se$_3$ Orthorhombic Metastable Phase

Paloma B. Souza[1], Milton A. Tumelero[2*], Ricardo Faccio[3], Rasin Ahmed[4], Cristiani C. Plá Cid[1], Giovanni Zangari[5], & Andre A. Pasa[1]

[1] Departamento de Física, Universidade Federal de Santa Catarina, 88.040-900 Florianópolis, Brazil

[2] Instituto de Física, Universidade Federal do Rio Grande do Sul, Caixa Postal 15051, 91501-970 Porto Alegre, Brazil

[3] Centro NanoMat & Área Física, DETEMA, Facultad de Química, Universidad de la República (Udelar), Montevideo C.P. 11800, Uruguay.

[4] Department of Electrical Engineering, University of Virginia, Charlottesville, Virginia 22904, USA

[5] Department of Materials Science and Engineering, University of Virginia, Charlottesville, Virginia 22904, USA

**Chalcogenides materials are being considered as some of the most promising systems for energy harvesting and energy conversion. Among them, the orthorhombic family of compounds X$_2$Y$_3$ (with X= Bi, Sb and Y= S, Se) has attracted special attention due to its interesting atomic structure, thermoelectric and optical properties. While Bi$_2$S$_3$ and Sb$_2$Se$_3$ have being already applied to solar cells, the new metastable Bi$_2$Se$_3$ is still a challenge due to the lack of data and knowledge on its properties. Here, the vibrational and structural properties of the orthorhombic metastable phase of Bi$_2$Se$_3$ were investigated by using Raman spectroscopy and ab-initio calculations. We perform Raman spectrum measurements along in-situ thermal treatment on samples grown by electrochemical deposition. We show that by increasing the temperature occurs an improved crystallization in the orthorhombic structure, followed by a recrystallization to the usual rhombohedral phase. Our results point out for specific Raman modes of the orthorhombic phase. First principle computational results based on density functional theory support the experimental data and describe three singlet Raman active vibrational modes, A$_g^1$, B$_{2g}$ and A$_g^2$.**

---

[*] Corresponding Authors: Milton A. Tumelero (matumelero@if.ufrgs.br) and Andre A. Pasa (aapasa@gmail.com).

**Introduction**

Bismuth selenide, $Bi_2Se_3$, is among the most studied materials during the last decade. The reason is that it may be the best so far known topological insulator [1,2], exhibiting among its benefits a facile synthesis, low toxicity, room temperature (RT) stability and, most importantly, a direct bandgap as large as 0.35 eV that could allow application up to room temperature.[3] Even with all this advantages, this material is still a challenge due to electrical issues[4] that prevent its utilization in electronics, spintronics and quantum computation. In the course of development for this material, during this last decade, several new proposals of applications for $Bi_2Se_3$ have been often discussed, such as photovoltaics[5], energy storage[6] and thermoelectric technology for refrigerators and thermogenerators[7]. The proposal of application in solar energy came with the possibility to synthesize a metastable crystalline phase with an orthorhombic structure[8,9], and space group *Pnma* that shows a direct bandgap energy of about 1.2 eV [8,10,11].

Several reports on solar cells using orthorhombic $Bi_2S_3$ and $Sb_2Se_3$ as sensitizer have shown excellent results[12,13]; these two materials in fact show a direct bandgap of 1.42 and 1.27 eV, respectively, which are close to the maximum attainable in the Shockley-Meissner limit[14]. The high open-circuit voltage observed is usually related to the excellent structural properties. The same materials were highlighted due to their superior thermoelectric properties, including the high Seebeck coefficient and the low thermal conductivity, reaching figures of merit of up to 0.6.[15] While orthorhombic $Bi_2Se_3$ is promising, the limited number of reports on its properties place this material among many of the possible chalcogenides. However, recent studies show that the bandgap of $Bi_2Se_3$ is even closer to the maximum attainable efficiency and can be used for the construction of tandem cells[10] or even thermoelectric applications[16].

The orthorhombic phase of $Bi_2Se_3$ is an isomorph of $Bi_2S_3$. Each bismuth is bonded to 6 selenium atoms in three distinct sites. The atoms are tightly bonded along the bc plane, but weakly grouped along the a-axis, giving a quasi-2D structure like the $Sb_2Se_3$ phase[12]. Such atomic arrangement propitiates the existence of inert surfaces, which have been reported as an outstanding property for solar cells, considerably reducing the dangling bonds and surfaces defects. Distinct from rhombohedral $Bi_2Se_3$, there is no van der Waals gap in between bc planes along the a-axis, nonetheless. This phase has been described as metastable, once it can be recrystallized to *R-3m* structure at relatively low annealing temperatures[17]. Experimentally, thin films of orthorhombic $Bi_2Se_3$ have been only grown by chemical[11] or electrochemical[8] synthesis at RT or low temperature. The reason is still unknown, but may be related to either the substrate[17] or electrodeposition parameters[9]. An alternative way to obtain the *Pnma* phase is to press the samples under high pressure[18], which keeps the orthorhombic phase under normal pressure conditions.

Despite the recent progress on synthesis of $Bi_2Se_3$, there is not much information available about its characteristics, such as vibrational properties. The opposite occurs instead for the rhombohedral *R-3m* phase, about which a significant amount of information is available. In this last case, the vibrational properties have been studied extensively, mainly focusing on the details and nature of the first few layers[19]. Four Raman shifts describe the vibrational spectrum of this material: two at frequencies about 39 and 75 cm$^{-1}$ due to active optical phonons with symmetry $E_g^{(1)}$ and $A_{1g}^{(1)}$, respectively; one other at about 133 cm$^{-1}$ with symmetry $E_g^{(2)}$ and a last one ($A_{1g}^{(2)}$) at about 175 cm$^{-1}$.[20] Another four Raman modes have been reported at the surface[21,22], such as an $A_1^{(3)}$ mode at about 160 cm$^{-1}$, usually observed in few quintuple layers[19] or superlattices[20]. A recent report of Raman spectroscopy of bulk samples that have a mixture of orthorhombic and rhombohedral phases shows the presence of a peak at about 157 cm$^{-1}$ which could be probably related to the orthorhombic phase and not to a rhombohedral surface mode[9].

Here, we present a structural and vibrational investigation of the orthorhombic phase of $Bi_2Se_3$ and compare it to the well-known rhombohedral phase. Theoretical results, based on ab initio calculations, indicate a rather complex vibration spectrum, showing a strong $A_g$ singlet mode which rules the Raman characterization. This $A_g$ mode was observed in orthorhombic $Bi_2Se_3$ samples grown by electrodeposition along with the Raman peaks of rhombohedral $Bi_2Se_3$ when thermal treatment is performed. Finally, the vibrational spectra were used to characterize the crystalline phase transition from orthorhombic to rhombohedral at about 140°C.

**Methods**

The samples used in this work are thin films of $Bi_2Se_3$ obtained by potentiostatic electrodeposition, with thickness of about 1 µm; preparation and overall characterization are described elsewhere[8,9,17]. The thermal annealing was carried out in a tubular oven, with a controlled argon atmosphere. Raman spectra were collected with a Renishaw In Via Reflex Micro Raman Spectrometer with laser light of wavelength $\lambda$ = 512 nm at 5% of laser power through a L50 objective and grating of 3000 l/mm. Also, Raman spectra were recorded during in-situ annealing of the sample. This experiment was performed in a Hot/Cold Stage filled with nitrogen gas. The temperature was increased from room temperature up to 540 K, in steps of 30 K, and at the end of each step one spectrum was recorded.

Geometrical optimizations were performed using *ab initio* methods within the Density Functional Theory (DFT) [23,24]. The calculations were made utilizing the VASP code (Vienna ab initio Simulation Package) [25–29]. The projector augmented wave (PAW) [29,30] method was employed to account for the electron–ion core interaction, utilizing the GGA (Generalized Gradient Approximation) exchange-correlation functional with the PBE parametrization (Perdew-Burke-Ernzerhof)[31]. PAW-PBE potentials were selected for Bi and Se atoms. The k-space grid was sampled until reaching a reasonably convergence. The cut-off energy for the plane wave expansion were selected as 500 eV. Structures were optimized until the forces in all the atoms were lower than 0.01 eV/Å.

Once the fully optimized unit cell and atomic positions were obtained, the determination of the vibrational information was carried out by means of Density Functional Perturbation Theory

(DFPT) [32–34], followed by a post-treatment by the Phonopy code[35,36]. Raman off-resonant activity spectra were calculated applying the methodology established by Porezag et al. [37] as implemented in a Python's script [38], with the objective of determining the polarizability derivate of every single phonon eigenmode at Γ-point. The structural models for the calculations were constructed based on the Rietveld refinement cell. The final optimized unit cell parameters corresponded to: $a = 4.183\text{Å}$ and $c = 29.454\text{Å}$ for the rhombohedral case; and $12.260\text{Å}$, $b = 4.147\text{Å}$ and $a = 11.608\text{Å}$ for the orthorhombic crystalline system.

**Results**

We start discussing the vibrational modes of *orthorhombic* phases of $Bi_2Se_3$ with ab initio calculations in the framework of DFT, as described in the methods section. The cell has a *Pnma* space group with $D_{2h}$ symmetry. The lattice parameters used in the calculations are slightly larger than the experimental ones, as expected from GGA functional, as displayed in Table I. The lattice parameter of *R-3m* phase is also presented in Table I. For $D_{2h}$ point group, eight 1D irreducible representation can be found, and about 60 phonon modes for a cell with 20 atoms, although, only 27 are Raman actives. The Raman active modes are the $A_g \oplus B_g \oplus B_{2g} \oplus B_{3g}$. The calculated phonon branches and density of states for both, *rhombohedral* and *orthorhombic* phases, along with a representation of atomic structure are presented in Figure 1(a-d). While the *rhombohedral* structure is composed by 12 transversal branches forming four large bands of phonons, seen in the density of vibrational states, the *orthorhombic* phase has many more branches and a large disperse phonon diagram.

The atomic arrangement of *R-3m* cell is well known, showing the quintuple layer (QL) composed by Se-Bi-Se-Bi-Se and interacting by van der Waals bonding leading to a large vacuum gap in the structure. The atomic structure of *Pnma* phase is rather complex. This form closed bilayer of $Bi_2Se_3$, which can be seen vertically in the Figure 1(b), with a certain atomic roughness. Such bilayers are weakly bonded to each other, but do not present vacuum gaps as the other phase. The importance of such closed bilayers have been previously discussed in the scope of $Sb_2Se_3$[12].

As discussed above, for the R-3m phase only 4 Raman active mode are expected among the 12 phonons branch of this crystal. The active Raman shifts and their respective irreducible representations are presented in Table II, for both phases of $Bi_2Se_3$, with comparison to experimental results (that will be presented in detail below) and calculation done elsewhere. Only the most active Raman modes are displayed. In the supplementary material all Raman modes can be found. The 13 most active Raman modes of *Pnma* phase are $8A_g \oplus 1B_g \oplus 2B_{2g} \oplus 2B_{3g}$. The $A_g^{(6)}$ at about 155.2 cm$^{-1}$ and the $B_{3g}^{(2)}$ at 123.4 cm$^{-1}$ are the ones with high intensity.

In order to compare the activities of the main Raman modes, the simulated spectrum of *orthorhombic* and *rhombohedral,* are presented in Figure 2. The simulation for the *R-3m* structure counts only with the $A_{1g}^{(2)}$ and $E_g^{(2)}$ modes and is in agreement with the literature.[39] The simulation for the *Pnma* structure counts with the 13 modes listed in the Table II.

The $A_g^{(6)}$ mode at about 155.2 cm$^{-1}$ is the dominant peak in the Bi$_2$Se$_3$ *orthorhombic* spectrum at Figure 2. The displacement vectors of the vibration are also presented in Figure 2, and show that this mode is due to displacement along the "*a*" direction of cell. This vibration is a stretching of the bilayer, discussed above, generating a kind of oscillation in the bilayer thickness. Previous results reported for the isomorphic Bi$_2$S$_3$, as a matter of comparison, the $A_{1g}^{(1)}$ and $B_{3g}^{(2)}$ mode were found at much higher frequencies 224.1 and 254.5 cm$^{-1}$, respectively[40]. This shift to higher frequencies could be related to a possible increase of bond order and energy, which also lead to smaller lattice parameter than Bi$_2$Se$_3$.

It is also important to mention that there is some peaks overlap on the spectrum of the two phases, once they have modes at almost same wavenumbers, such as at about 130 cm$^{-1}$, where the *R-3m* phase shows the $E_g^{(2)}$ mode and the *Pnma* has the active $B_{3g}^{(2)}$ mode. At about 173 cm$^{-1}$, the *orthorhombic* $B_{2g}^{(2)}$ peak is also close in energy to *rhombohedral* $A_{1g}^{(2)}$. We notice, however, that the intensity on these peaks is distinct in both spectra in Figure 2. The $E_g^{(2)}$ is more intense than $A_{1g}^{(2)}$ in the *R-3m* case, however, for the *Pnma* case the $B_{2g}^{(2)}$ shows higher activity than $B_{3g}^{(2)}$. Such behavior could be useful to separate contributions of both phases.

The Raman spectrum from the rhombohedral phase is widely reported and is in good agreement to the theoretical data described above. The Raman modes for the *orthorhombic* phase, nonetheless, to the best of our knowledge, have not been presented so far. Here, to proceed with the investigation of vibrational properties of *Pnma* structure we are carrying out a complete evaluation of Raman spectrum of samples presenting *orthorhombic* crystalline phase.

The samples used in this work present, in majority, *orthorhombic* (*Pnma*) phase, with a small mixture of nanocrystalline *rhombohedral* grains. For more information about samples preparation refer to methods section and Ref. 9. On previous reports two intense and characteristic peaks at about 128.4 and 173.1 cm$^{-1}$ of *R-3m* phase of Bi$_2$Se$_3$ were clearly observed, due to the $E_g^{(2)}$ and $A_{1g}^{(2)}$, respectively. [9,17] A third peak at about 158.7 cm$^{-1}$ could also be seen in these reports, but with very small intensity. By adding 0.06 mM CuCl$_2$ to the preparation electrolyte of references 8 and 9, the two peaks from *R-3m* phase, described above are strongly reduced, making that at 158.7 cm$^{-1}$ more relevant. These changes in Raman spectrum could be taken as an evidence of the increase of *Pnma* phase over *R-3m* in samples produced with CuCl$_2$ additive. The XRD (not shown) did not indicate changes, nonetheless.

The Raman spectra measured at different temperatures of a sample prepared with the addition of 0.06 mM CuCl$_2$ to the deposition electrolyte is presented in Figure 3(a), from RT to 540 K. In the Figure 3(b) is presented the spectrum at 330 K, which shows better the peak from *orthorhombic* phase. Other than the peak at 158.7 cm$^{-1}$, four other peaks could be identified by using Lorentzian peak profiling. All peaks in the fitting procedure have almost the same FWHM of about 20 cm$^{-1}$, which is large compared to single crystal, but acceptable for polycrystalline thin films. Five fitting peaks were used to better adjust the Raman data. The following peaks were observed 109.0, 126.6, 141.1, 158.7 and 173.9 cm$^{-1}$. They were attributed to the modes $A_g^{(4)}$, $B_{3g}^{(2)}$, $B_{2g}^{(1)}$, $A_g^{(6)}$ and $B_{2g}^{(2)}$. No comparison was found in literature. The peak at 158.9 cm$^{-1}$ in as-grown samples of Bi$_2$Se$_3$ has been previously discussed as a possible surface mode of *R-3m* Bi$_2$Se$_3$ with $A^{(3)}$ symmetry[9]. Here, by using DFT analysis we show that such peak is due to the orthorhombic phase vibrational mode with $A_g$ symmetry, assigned as $A_g^{(6)}$.

Increasing the temperature, from RT to 330 K, the peaks in the spectrum become narrower, which is attributed to improvements in the *Pnma* crystalline properties due to the low temperature annealing. At temperature of 360 K, the peaks $A_{1g}^{(2)}$ and $E_g^{(2)}$ from *rhombohedral* Bi$_2$Se$_3$ start to appear at about 132.0 and 176.2 cm$^{-1}$, respectively. By further increasing the temperature one can see that $A_{1g}^{(2)}$ and $E_g^{(2)}$ modes become more intense while the peak $A_g^{(6)}$ at about 158.7 cm$^{-1}$ decreases and vanishes at 510 K. Clearly, above 360 K and up to 480 K both crystalline phases coexist in the Raman spectrum. At temperatures of 510 and 540 K, only the rhombohedral modes are observed. Figure 3(c) displays the Raman spectrum at 540 K, showing clearly the $A_{1g}^{(2)}$ and $E_g^{(2)}$ modes at frequencies 130.3 and 174.2 cm$^{-1}$, respectively. The *Pnma* phase is no longer detected in this samples. Similar result on recrystallization by heating was previously discussed on the basis of X-Ray Diffraction analysis.[9]

Figure 3(d) shows the position of the peaks as a function of temperature obtained by fitting the peak profile with Lorentzian functions. All peaks present a frequency red-shift by increasing the temperature, which is probably related to the increase of lattice parameter due to thermal expansion, as discussed in Refs. 40 and 41. The rate of frequency decrease of the Raman modes of rhombohedral phase have been properly described by Irfan et al.[41] with a linear behavior for the frequency-temperature relation of the modes $A_{1g}^{(1)}$, $E_g^{(2)}$ and $A_{1g}^{(2)}$, with coefficients of -0.0194 cm$^{-1}$/K for the $E_g^{(1)}$ and -0.0195 cm$^{-1}$/K for the $A_{1g}^{(2)}$, at temperatures ranging from 83 to 523 K. Kim et al.[42], found a rather small value for the temperature coefficient of $A_{1g}^{(2)}$ mode, about -0.015 cm$^{-1}$/K, showing that the thermal expansion accounts for 40% of the change in frequency in single crystals.

Here, our results from 300 K to 570 K give for $A_{1g}^{(2)}$ a coefficient of -0.0125 cm$^{-1}$/K, smaller than previous reports. Additionally, our results point out to small changes in frequency of rhombohedral $E_g^{(2)}$ and orthorhombic $A_g^{(6)}$ modes. The reason could be intimately related to the overall crystalline structure. While previous reports were based on observations in single crystals, here, we describe results from polycrystalline thin films. The dashed lines in the graph of Figure 2(d) are the linear fits that give the temperature coefficients. The continuous lines in the data for the $E_g^{(2)}$ and $A_{1g}^{(2)}$ modes stand for the simulation of Raman shift changes based only on thermal expansion, using the equation below:

$$f(T) = f_0 e^{-\gamma T(2\alpha_a + \alpha_c)}, \qquad \text{Eq. 1}$$

where $f_0$ is the zeroth-temperature Raman shift. The equation was adapted from Ref. [42], due to quasi-harmonic corrections considering the thermal expansion of the lattice. The expansion coefficients were assumed to be almost constant in the annealing range and equal to 1.9 x10$^{-5}$ K$^{-1}$ and 1.1 x 10$^{-5}$ K$^{-1}$ for $\alpha_a$ and $\alpha_c$, respectively.[43] The $\gamma$ is the Grüneisen parameter for the Bi$_2$Se$_3$ equal to 1.4.[42] For the simulation, the $f_0$ constants for each mode E$_g^{(2)}$ and A$_{1g}^{(2)}$ were obtained from extrapolation of linear fit, which gives 133.2 and 180.1 cm$^{-1}$, respectively. Notice that the simulation fairly agrees with the A$_{1g}^{(2)}$ data, indicating that the thermal expansion may be the main origin of the red-shift in this mode. The simulation for the E$_g^{(2)}$ indicates that much higher changes should be expected. No simulation was done for the peak at 158.7 cm$^{-1}$ once it is being attributed to the *orthorhombic* phase of Bi$_2$Se$_3$, and no estimation of thermal expansion or Grüneisen parameter could be found in the literature.

Based on the results described above, we argue that in these thin films, the thermal expansion effects on the Raman shifts could be smaller than in the single crystal, since the silicon substrate has a much small expansion coefficient than Bi$_2$Se$_3$, in this case, the substrate generates strain in the lattice, which could be responsible for the lack of red-shift in E$_g^{(2)}$ mode.

At this point, we consider that the peaks at about 132.0 and 176.2 cm$^{-1}$ should better addressed. Note that these peaks appear in both spectra, in Figure 3(b) and 3(c), with just small deviations in frequency, both peaks are expected in these two phases, nonetheless, as shown in Table II. The peaks in Figure 3(b) were indexed as *orthorhombic* phase while the peaks in Figure 3(c) as *rhombohedral* ones. Our analysis was based on the relative peak intensities. Raman modes for the *R-3m* have been extensively reported with E$_g^{(2)}$ peak always higher in intensity than A$_{1g}^{(2)}$ [9,39,41,44], in agreement to DFT results. For the *Pnma* phase, however, the peak at 126.6 cm$^{-1}$ is less intense than the one at 173.9 cm$^{-1}$. The Raman data at low temperature follow the peak relation for the *orthorhombic* phase, while increasing the temperature, the peak profile changes and agrees with the expected for the *rhombohedral* phase.

As an example of the Raman peak indexation described above, we are presenting in Figure 4 Raman spectra of a sample highly crystalline orthorhombic phase electrodeposited with the electrolyte describe in Ref. 17, without the addition of CuCl$_2$, obtained with distinct excitation lasers. Four excitations were used, 785 nm (IR), 514 nm (green), 488 nm (blue) and 405 nm (violet). By using high energy excitations (violet, blue and green wavelengths), the peaks of *Pnma* phase at 126 cm$^{-1}$ and 158 cm$^{-1}$) are more intense. The dashed vertical lines in the figure clearly shows the Raman shift for the two phases. For the 405 nm laser, the Raman spectrum is the one closest to the expected for the *Pnma* based on computation results of Figure 2. The spectra with IR excitation however shows a more pronounced Raman shift for B$_{1g}^{(1)}$ and A$_g^{(8)}$ modes of the *Pnma* phase. Due to inversion symmetry of Bi$_2$Se$_3$ crystal the optical modes of *R-3m* phase have been reported to be either Raman active or IR-active.[45] The signal intensity for IR excitation was also enhanced by 20-fold compared to those obtained for the other excitation wavelengths. Glinka et al.[46] reported a 100-fold enhancement of the Raman signal by switching from 532 nm to 785 nm which was attributed in part to the optical coupling to surface states of the *R-3m* phase[46] The enhancement and emergence of IR-active modes when using IR laser could be due to an electronic resonant transitions at the 1.58 eV. Alternatively, the depth of penetration of IR laser all the way down to the interface between film and substrate could explain the emergence of 131 cm$^{-1}$ $E_g^{(2)}$ and 173.3 cm$^{-1}$ $A_{1g}^{(2)}$ modes of the *R-3m* phase since the previous results indicate the existence of a mixture of *R-3m* and *Pnma* phases during early stages of Bi$_2$Se$_3$ film growth (near the substrate) by electrodeposition[8]. The relation of laser wavelength and phase selectivity could be discussed as dependent on the penetration depth of the excitation, in accordance with the measurement with 405 nm laser, that is more sensitive to the surface layers.

In conclusion, the vibrational properties of the Bi$_2$Se$_3$ *orthorhombic* phase were calculated with DFT showing an intense peak for the $A_{1g}^{(6)}$ mode at about 155.2 cm$^{-1}$, which is in good agreement with experimental results. Other modes were also found in experimental data as well as in the computational results. To the best of our knowledge this is the first report on vibration and Raman properties of *Pnma* phase of Bi$_2$Se$_3$ compound. Additionally, the existence of an *orthorhombic* phase was confirmed in low-temperature-prepared Bi$_2$Se$_3$ samples. The thermal annealing of the samples promotes an improvement of the crystal lattice of *orthorhombic* phase from RT to 330 K, but above 360 K the *rhombohedral* phase starts to recrystallize from *orthorhombic*. At temperatures higher than 510 K all *orthorhombic* phase was totally transformed in *rhombohedral*. The low temperature threshold for recrystallization of this phase is attributed to a metastable equilibrium, where the *Pnma* phase occurs in a local shallow energy minimum.

**Acknowledgments**

The authors would like to acknowledge the funding agencies CNPQ and CAPES. MAT acknowledge the grants CNPQ 310171/2017-2 and project FAPERGS ARD/2017-0820. RF acknowledges PEDECIBA, CSIC-Udelar and ANII, Uruguayan Institutions. The authors also thank the CENAPAD-SP for the computational resources

Table I: Lattice parameters obtained by DFT calculation in comparison to the values found experimentally.

|  |  | a (Å) | b (Å) | c (Å) |
|---|---|---|---|---|
| R-3m | Experimental | 4.135 | - | 28.765 |
|  | Theoretical | 4.183 | - | 29.454 |
| Pnma | Experimental | 11.710 | 4.140 | 11.430 |
|  | Theoretical | 12.260 | 4.147 | 11.608 |

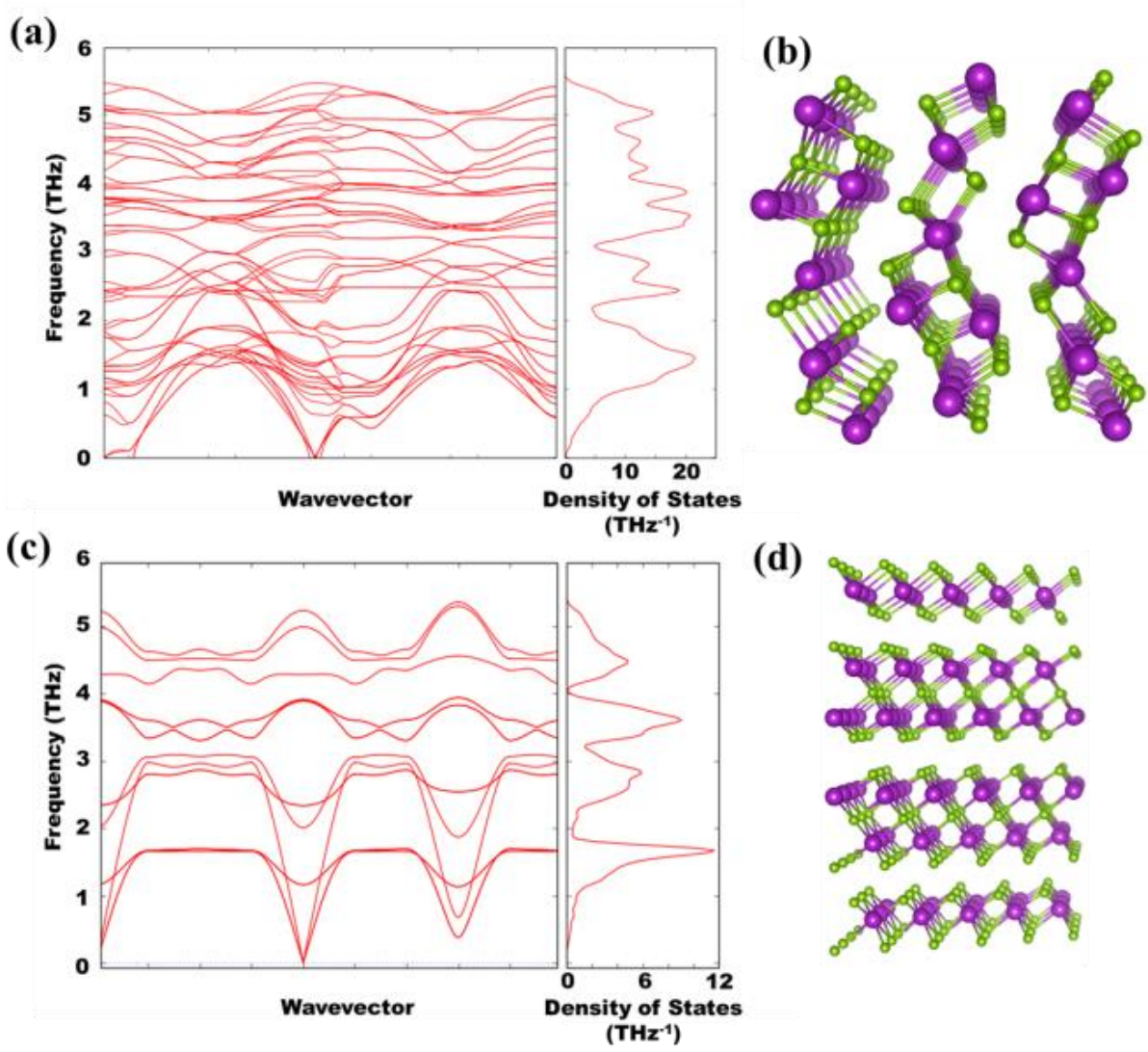

**Figure 1: Phonon dispersion and phonon density for the (a) *Pnma* and (c) *R-3m* phases of $Bi_2Se_3$. (b) and (d) are the atomic structures of orthorhombic and rhombohedral phases, respectively. Images processed with VESTA[47].**

**Table II: Raman active modes for the *R-3m* and *Pnma* phases of Bi$_2$Se$_3$ and its respective frequencies, calculated with DFPT, measured in this work and in comparison to previous reports.**

|  | Rhombohedral (*R-3m*) Raman Modes (cm$^{-1}$) | | |
|---|---|---|---|
|  | Theoretical | Experimental | Literature |
| E$_g^{(1)}$ | 38.7 | -- | 37 [41] |
| A$_{1g}^{(1)}$ | 66.9 | -- | 72 [41] |
| E$_g^{(2)}$ | 130.4 | 132 | 131 [41] |
| A$_{1g}^{(2)}$ | 174.5 | 177 | 175 [41] |
|  | Orthorhombic (*Pnma*) Raman Modes (cm$^{-1}$) | | |
| B$_{3g}^{(1)}$ | 36.4 | -- | -- |
| A$_g^{(1)}$ | 45.9 | -- | -- |
| A$_g^{(2)}$ | 81.5 | -- | -- |
| A$_g^{(3)}$ | 108.1 | 109.0 | -- |
| A$_g^{(4)}$ | 112.9 |  | -- |
| B$_{3g}^{(2)}$ | 123.4 | 126.6 | -- |
| B$_{1g}^{(1)}$ | 130.7 | -- | -- |
| A$_g^{(5)}$ | 137.1 | 141.1 | -- |
| B$_{2g}^{(1)}$ | 147.9 |  | -- |
| A$_g^{(6)}$ | 155.2 | 158.7 | -- |
| A$_g^{(7)}$ | 160.4 | -- | -- |
| B$_{2g}^{(2)}$ | 169.2 | 173.0 | -- |
| A$_g^{(8)}$ | 171.6 |  | -- |

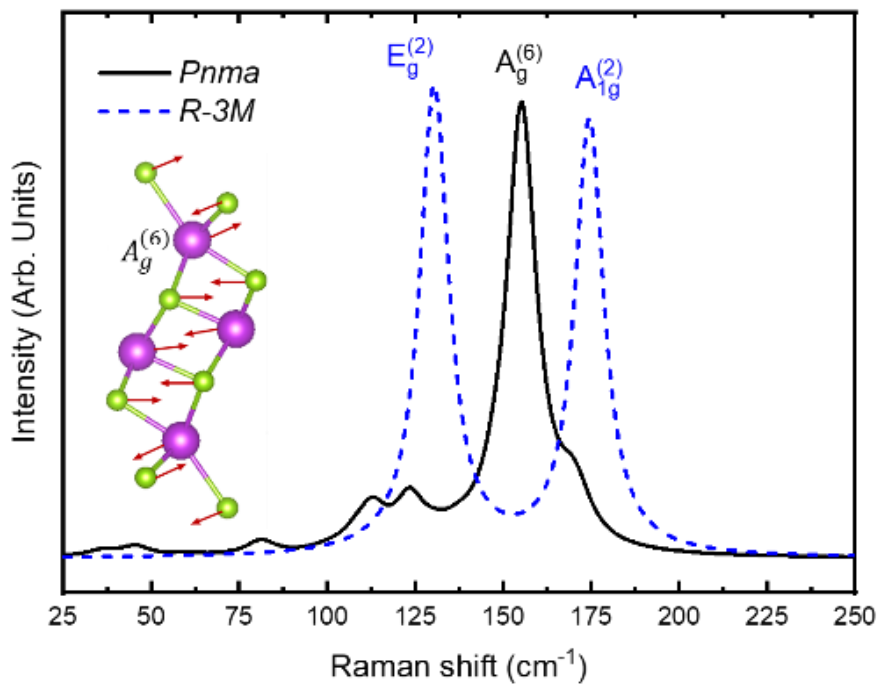

**Figure 2:** Simulated Raman spectra for both *Pnma* and *R-3m* crystallographic structure of $Bi_2Se_3$. The inset with the atomic scheme shows the displacement vectors for the $A_g^{(6)}$ mode of *Pnma* phase.

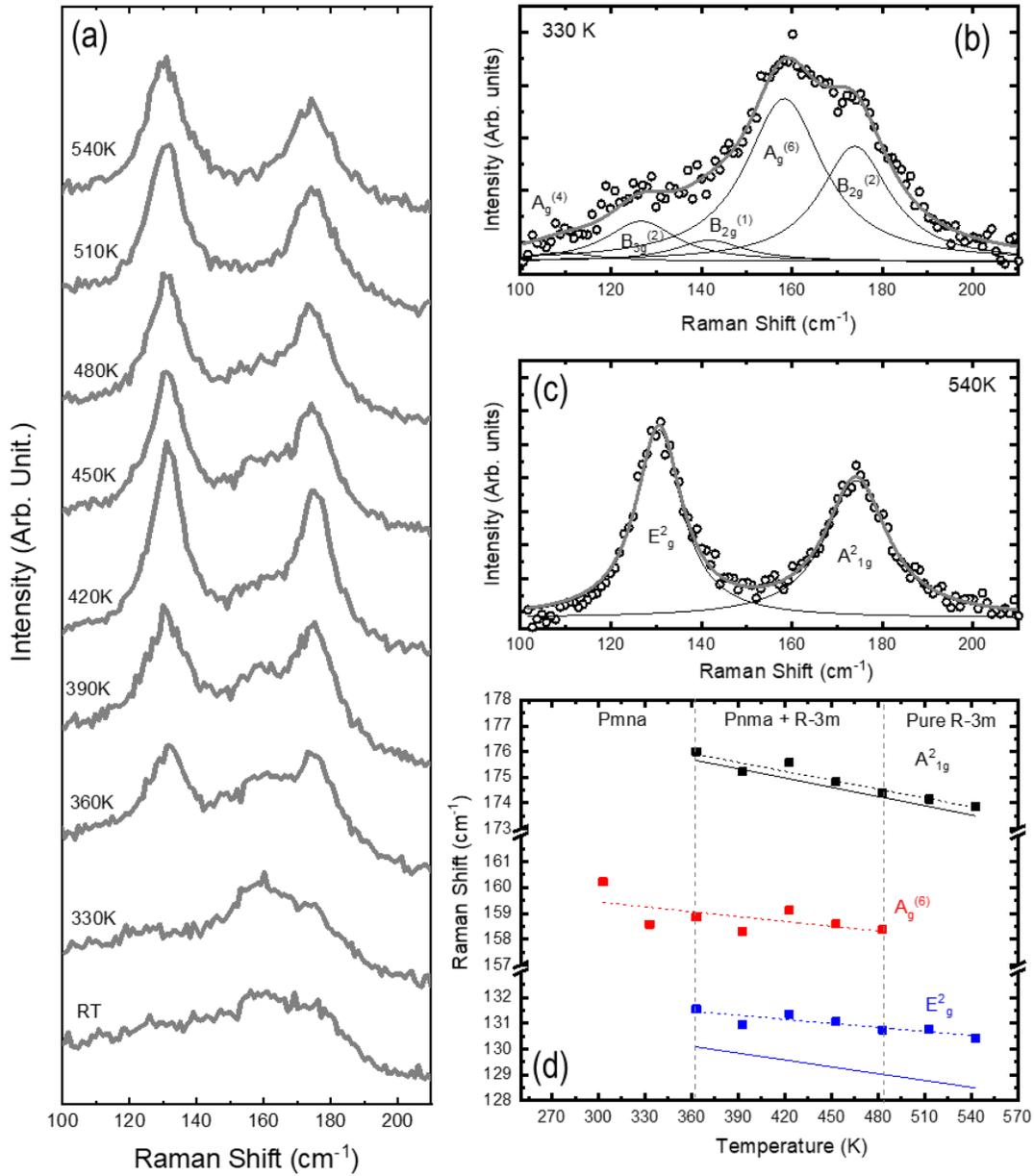

**Figure 3:** In (a) Raman spectra at different temperatures. (b) and (c) fitted Raman spectra for temperatures of 330 and 540 K, respectively, and (d) peak position of $E_g^{(2)}$, $A_g^{(6)}$, and $A_{1g}^{(2)}$ as a function of temperature, where the continuous lines are the linear fit and the dashed line the simulated curves.

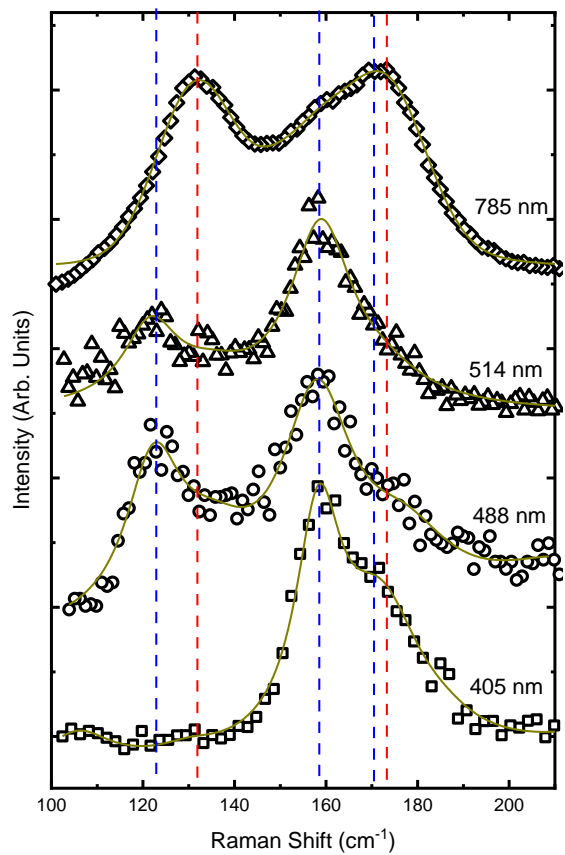

**Figure 4: Raman spectrum for a sample measured with distinct laser excitation** [2]**. The blue vertical dashed line indicates the peak position expected for the *Pnma* phase, the red dashed line indicates the position of *R-3m* Raman modes.**